\documentclass[preprint,12pt]{elsarticle}

\usepackage{graphics}

\usepackage{amsmath}	
\usepackage{amssymb}    
\usepackage{bbm}
\newcommand{\be}{\begin{eqnarray}}
 \newcommand{\ee}{\end{eqnarray}}

\newcommand{\boldnabla}{\mbox{\boldmath$\nabla$}}

\usepackage{bibentry}

\newcommand{\ignore}[1]{}
\newcommand{\nobibentry}[1]{{\let\nocite\ignore\bibentry{#1}}}

\journal{Computer Physics Communications} 

\begin{document}

\begin{frontmatter}



\title{Numerical Solution of the Time-Dependent Dirac Equation in Coordinate Space without Fermion-Doubling}

\author[crm,carl,fields]{Fran\c{c}ois Fillion-Gourdeau}
\ead{filliong@CRM.UMontreal.ca}

\author[carl,crm]{Emmanuel Lorin}
\ead{elorin@math.carleton.ca}

\author[sher,crm]{Andr\'{e} D. Bandrauk}
\ead{andre.bandrauk@usherbrooke.ca}


\address[crm]{Centre de Recherches Math\'{e}matiques, Universit\'{e} de Montr\'{e}al, Montr\'{e}al, Canada, H3T~1J4}
\address[carl]{School of Mathematics and Statistics, Carleton University, Ottawa, Canada, K1S 5B6}
\address[sher]{Laboratoire de chimie th\'{e}orique, Facult\'{e} des Sciences, Universit\'{e} de Sherbrooke, Sherbrooke, Canada, J1K 2R1}
\address[fields]{Fields Institute, University of Toronto, Toronto, Canada, M5T 3J1}

\date{\today}
\begin{abstract}
The validation and parallel implementation of a numerical method for the solution of the time-dependent Dirac equation is presented. This numerical method is based on a split operator scheme where the space-time dependence is computed in coordinate space using the method of characteristics. Thus, most of the steps in the splitting are calculated exactly, making for a very efficient and unconditionally stable method. We show that it is free from spurious solutions related to the fermion-doubling problem and that it can be parallelized very efficiently. We consider a few simple physical systems such as the time evolution of Gaussian wave packets and the Klein paradox. The numerical results obtained are compared to analytical formulas for the validation of the method. 
\end{abstract}


\end{frontmatter}





\section{Introduction}

The Dirac equation is very important in many fields of physics and chemistry because it gives a relativistic description of electrons and other spin-$\frac{1}{2}$ particles. For this reason, it has been studied and used extensively in relativistic heavy ion collisions, heavy ion spectroscopy and more recently, in laser-matter interaction (for a review, see \cite{Salamin200641} and references therein) and condensed matter physics \cite{Katsnelson2006}. However, solving this equation is still a very challenging problem even if it has been derived more than 80 years ago and has been utilized profusely. The number of closed-form solutions is very limited due to the intricate structure of Dirac matrices which couple the components of the four-spinor wave function. For this reason, only highly symmetric systems can be studied by analytical means; the mathematical description of more realistic systems should be based on approximation methods such as semi-classical theory \cite{PhysRevLett.89.193001} or numerical calculations. However, the typical time scale of the electron dynamics is usually much smaller than the time scales of interesting phenomena, rendering the numerical solution notoriously difficult and requiring a lot of computer resources. Moreover, certain numerical techniques (such as the naive symmetric difference scheme) are plagued with spurious solutions related to the fermion-doubling problem \cite{PhysRevD.26.468,Muller1998245,Kogut:1974ag}. Of course, such issues should be addressed by any numerical methods to obtain accurate solutions describing physically relevant systems.

The most popular numerical methods rely usually on some variations of the split operator method along with a spectral scheme \cite{PhysRevA.59.604,Mocken2008868,Mocken2004558,PhysRevA.53.1605,Bauke2011}. These approaches have also been used to solve the coupled Maxwell-Dirac system of equations \cite{Huang2005761,Bao2004663}. In real space, without resorting to a spectral method, the finite element scheme \cite{Muller1998245,PhysRevLett.54.669} and the finite difference scheme (both explicit \cite{0022-3700-16-11-017} and implicit \cite{PhysRevA.40.5548,PhysRevA.79.043418,PhysRevC.71.024904}) have also been exploited. A major complication is usually shared by these ``real space'' approaches: they often exhibit spurious states related to the fermion doubling problem \cite{PhysRevD.26.468,RevModPhys.55.775}. The latter leads to the appearance of new unphysical modes when the Dirac equation is discretized and is an artifact of the discretization process (a more thorough discussion of this subject will be presented in Section \ref{sec:fermi_double}). To solve this problem, some variations of the Wilson \cite{RevModPhys.55.775} and staggered \cite{RevModPhys.55.775,PhysRevD.16.3031,PhysRevD.13.1043,PhysRevD.11.395} fermion discretizations have been used. The latter were formulated to solve the fermion doubling problem in gauge field theory on the lattice. The main issue with these methods is twofold. First, they usually break chiral symmetry, one of the fundamental symmetries of the Dirac equation. Second, from the numerical point of view, they often lead either to diffusive, unstable or non-conservative (probability conservation) schemes.

In this work, we are implementing and analyzing a variation of the ``real space'' method presented in \cite{Lorin2011190} and show that it circumvents some of the issues described previously. This numerical method is a combination of the split operator technique and of the method of characteristics; the latter is used to obtain 1-D analytical solutions of the Dirac equation in each dimension. These solutions are utilized to evolve the wave function at every time step by alternate direction iteration and thus, the time-evolution is exact for almost every step of the operator splitting (by choosing carefully the time and space increments). The resulting scheme is very similar to an upwind scheme with a Courant-Friedrichs-Lewy condition CFL = 1. Because exact solutions are used in the time evolution, there is no fermion doubling and most symmetries are preserved (such as chiral symmetry). Moreover, it yields a very simple numerical method which is unconditionally stable and which can be efficiently parallelized.  

This article is separated as follows. In Section \ref{sec:num_meth}, the numerical method and the discretization of the Dirac equation is presented. In Section \ref{sec:fermi_double}, a description of the fermion-doubling problem is exposed along with a proof that the numerical scheme is free from these unphysical states. Some details of the implementation and code performance, such as the parallelization efficiency, are shown in Section \ref{sec:perf}. In Section \ref{sec:validation}, some simple physical systems are analyzed to demonstrate the validity of our numerical approach. The time evolution of a Gaussian wave packet and the Klein paradox are evaluated and compared to analytical results. We conclude in Section \ref{sec:conclu}.

\section{Numerical Methods}
\label{sec:num_meth}

The Dirac equation describes the relativistic dynamics of spin-$\frac{1}{2}$ particles (fermions) such as electrons and quarks. By construction, it is a one-particle theory which is relativistically covariant, i.e. it is invariant under Poincar\'{e} transformations. Other relativistic generalization of the Schr\"odinger equation also exists, such as the Klein-Gordon equation for example. These wave functions are distinguished mostly by their spin content (and possibly other quantum numbers). For instance, it can be shown, by looking at Lorentz transformations, that the Klein-Gordon equation describes spin-0 particles while the Dirac equation describes spin-$\frac{1}{2}$ particles \cite{Itzykson:1980rh}. Of course, the right choice of wave equation depends on the system under investigation. 

In this work, we are interested by the relativistic dynamics of an electron of mass $m$ coupled to an external classical electromagnetic field characterized by its electromagnetic potential. The latter is introduced in the Dirac wave equation via the minimal coupling prescription\footnote{The minimal coupling prescription consists in replacing $\partial_{\mu} \rightarrow \partial_{\mu} + ie A_{\mu}$, where $A_{\mu}$ is the electromagnetic potential with Lorentz index $\mu$.} which ensures the invariance of the resulting equation under gauge transformations \cite{Itzykson:1980rh}. Therefore, the time-dependent Dirac equation is given by \cite{Itzykson:1980rh}
\begin{eqnarray}
i\partial_t \psi(t,\mathbf{x}) = \hat{H} \psi(t,\mathbf{x})
\label{eq:dirac_eq}
\end{eqnarray}
where $\psi(t,\mathbf{x})$ is the time and coordinate dependent four-spinor and $\hat{H}$ is the Hamiltonian operator. The latter is given by
\begin{eqnarray}
\hat{H}  =  \boldsymbol{\alpha} \cdot \left[  c\mathbf{p} - e\mathbf{A}(t,\mathbf{x}) \right] + \beta m c^{2} + e\mathbb{I}_{4}V(t,\mathbf{x}) .
\label{eq:hamiltonian}
\end{eqnarray}
where the momentum operator is $\mathbf{p} = -i \boldnabla$. Here, $\mathbf{A}(t,\mathbf{x})$ represents the three space components of the electromagnetic vector potential, $V(t,\mathbf{x}) = A_{0}(t,\mathbf{x})$ is the scalar potential, $e$ is the electric charge (obeying $e=-|e|$ for an electron), $\mathbb{I}_{4}$ is the 4 by 4 unit matrix and $\boldsymbol{\alpha}=(\alpha_i)_i,\beta$ are the Dirac matrices. In all calculations, the Dirac representation is used where
\begin{eqnarray}
\alpha_{i} = 
\begin{bmatrix}
	0 & \sigma_{i} \\
	\sigma_{i} & 0 
\end{bmatrix}
 \; \; , \; \;
\beta = 
\begin{bmatrix}
	\mathbb{I}_{2} & 0 \\
	0 & -\mathbb{I}_{2} 
\end{bmatrix} .
\label{eq:dirac_mat}
\end{eqnarray}
The $\sigma_{i}$ are the usual $2 \times 2$ Pauli matrices defined as
\begin{eqnarray}
\sigma_{x} = 
\begin{bmatrix}
0 & 1 \\ 1 & 0  
\end{bmatrix}
\;\; \mbox{,} \;\;
\sigma_{y} = 
\begin{bmatrix}
0 & -i \\ i & 0 
\end{bmatrix}
\;\; \mbox{and} \;\;
\sigma_{z} = 
\begin{bmatrix}
1 & 0 \\ 0 & -1 
\end{bmatrix},
\end{eqnarray}
while $\mathbb{I}_{2}$ is the 2 by 2 unit matrix. Note that the light velocity $c$ and fermion mass $m$ are kept explicit in Eq. (\ref{eq:hamiltonian}), allowing to adapt the method easily to natural or atomic units (a.u.). 

Throughout this work, we consider the single particle Dirac equation which is relevant for calculations describing Quantum Electrodynamics (QED) processes coupled to strong classical fields (e.g. particle-antiparticle pair creation from very high intensity electromagnetic classical fields). In these conditions, the negative energy states of the Dirac equation have a well-defined physical interpretation and have to be included. However, when QED processes become less important but relativistic effects still need to be taken into account, the negative energy states are filled, according to the Dirac sea picture. Thus, transitions to the negative energy spectrum are forbidden. Numerically, this physical requirement can be implemented by adding projection operators at appropriate places in the calculation (see \cite{PhysRevA.79.043418} for a discussion of this issue), allowing to consider positive energy states only. This however is not investigated in this study, although in principle, it could be implemented in our numerical method.

%
%

\subsection{Operator Splitting (lowest order)}

The problem at hand is to find an approximate solution of Eqs. (\ref{eq:dirac_eq}) and (\ref{eq:hamiltonian}) at a certain time $t_{n+1}$ given by $\psi^{n+1}(\mathbf{x})$ with an initial condition at time $t_{n}$ given by
\begin{equation}
 \psi(t_{n},\mathbf{x}) = \psi^n (\mathbf{x}). 
\end{equation}
As previously discussed in \cite{Lorin2011190}, this can be done with an operator splitting scheme, which forms the basis of our numerical method. To be more specific, let us define the operators
\begin{eqnarray}
\label{eq:op_A}
\hat{A} &=& -i c\alpha_{x} \partial_{x} \\
\label{eq:op_B}
\hat{B} &=& -i c\alpha_{y} \partial_{y}  \\
\label{eq:op_C}
\hat{C} &=& -i c\alpha_{z} \partial_{z}  \\
\label{eq:op_D}
\hat{D} &=& \beta m c^{2} + e\mathbb{I}_{4}V(t,\mathbf{x}) - e\boldsymbol{\alpha} \cdot \mathbf{A}(t,\mathbf{x}).
\end{eqnarray}
The following splitting in Cartesian coordinates is considered \cite{Lorin2011190} (note that we omit the $\mathbf{x}$ in the wave functions argument for notational convenience):
\begin{eqnarray}
\label{eq:split1}
i\partial_{t} \psi^{(1)}(t) &=& \hat{A} \psi^{(1)}(t), \; \psi^{(1)}(t_{n}) = \psi^n,  \;\;\;\;\;\;\;\;\;\;\; t \in [t_{n},t_{n+1}) \\
\label{eq:split2}
i\partial_{t} \psi^{(2)}(t) &=&\hat{B} \psi^{(2)}(t), \; \psi^{(2)}(t_{n}) = \psi^{(1)}(t_{n+1}), \; t \in [t_{n},t_{n+1}) \\
\label{eq:split3}
i\partial_{t} \psi^{(3)}(t) &=& \hat{C} \psi^{(3)}(t), \; \psi^{(3)}(t_{n}) = \psi^{(2)}(t_{n+1}), \; t \in [t_{n},t_{n+1})  \\
\label{eq:split4}
i\partial_{t} \psi^{(4)}(t) &=& \hat{D} \psi^{(4)}(t), \; \psi^{(4)}(t_{n}) = \psi^{(3)}(t_{n+1}), \; t \in [t_{n},t_{n+1})  \\
\mbox{and} \;\; \psi^{n+1} &=& \psi^{(4)}(t_{n+1})
\end{eqnarray}
where the upper subscript in parenthesis on the wave function denotes the splitting step number. Note that this splitting scheme leads to an error that scales like ($O(\delta t^{2})$), leading to a first order numerical scheme (for more details on the analysis of the method, see \cite{Lorin2011190}). 
For each step, the initial condition and time domain are shown explicitly. Written like this, the method consists of solving each equation independently with an initial condition given by the solution of the previous step. Note also that for every step, the time increment is the same, i.e. $\delta t \equiv t_{n+1} - t_{n}$. This way of splitting the Dirac equation is very similar to the more usual ``T-exponential'' operator splitting approach used in \cite{PhysRevA.59.604,Mocken2008868,Bao2004663,Huang2005761}, as can be seen by computing the Fourier transform in space coordinates of Eqs. (\ref{eq:split1}) to (\ref{eq:split3}). These numerical techniques, using the evolution operator, are usually based on spectral methods which require the computation of the discrete Fourier transform at every time step. It is possible to circumvent the spectral methods altogether by noting that an analytical solutions of Eqs. (\ref{eq:split1}) to (\ref{eq:split4}) can be derived, as discussed in the following.

\subsection{Method of characteristics}

Eqs. (\ref{eq:split1}) to (\ref{eq:split3}) can be solved independently using the method of characteristics. This essentially proceeds in three steps. First, the Dirac matrix is transformed to a new representation where it is diagonalized, thus decoupling the spinor components. The resulting equation has a form similar to an advection or transport equation (linear first order in time and space derivative). The method of characteristics is then used to find an explicit analytical solution. Finally, this solution is transformed back to the original Dirac matrix representation. The explicit solution of these equations and computational details are relegated to  \ref{sec:app1}. The final result for the solutions of Eqs. (\ref{eq:split1}) to (\ref{eq:split3}) is 
\begin{eqnarray}
\label{eq:sol_ex_1}
 \psi^{(1)}(t_{n+1},\mathbf{x}) &=& \frac{1}{2} \biggl\{ [\mathbb{I}_{4} + \alpha_{x}] \psi^n(x-c\delta t,y,z)  \nonumber \\
&& \;\;\;+ [\mathbb{I}_{4} - \alpha_{x}] \psi^n(x+c \delta t,y,z)  \biggr\} \\
\label{eq:sol_ex_2}
 \psi^{(2)}(t_{n+1},\mathbf{x}) &=& \frac{1}{2} \biggl\{ [\mathbb{I}_{4} + \alpha_{y}] \psi^{(1)}(t_{n+1},x,y-c\delta t,z)  \nonumber \\
&&\;\;\;+ [\mathbb{I}_{4} - \alpha_{y}] \psi^{(1)}(t_{n+1},x,y+c \delta t,z)  \biggr\} \\
\label{eq:sol_ex_3}
 \psi^{(3)}(t_{n+1},\mathbf{x}) &=& \frac{1}{2} \biggl\{ [\mathbb{I}_{4} + \alpha_{z}] \psi^{(2)}(t_{n+1},x,y,z-c\delta t) \nonumber \\
&& \;\;\;+ [\mathbb{I}_{4} - \alpha_{z}] \psi^{(2)}(t_{n+1},x,y,z+c \delta t)  \biggr\} 
\end{eqnarray}
Each of these solutions represents traveling waves moving in opposite directions at velocity $c$. 

In the last equation of the splitting, Eq. (\ref{eq:split4}), the solution is simply given by
\begin{eqnarray}
\label{eq:sol_ex_4}
 \psi^{(4)}(t_{n+1},\mathbf{x}) &=& T \exp \left[ -i \int_{t_{n}}^{t_{n+1}} d\tau \left[ \beta mc^{2} 
-e \boldsymbol{\alpha} \cdot   \mathbf{A}(\tau,\mathbf{x}) \right]  \right] \nonumber \\
&& \times \exp \left[-ie \int_{t_{n}}^{t_{n+1}} d\tau V(\tau,\mathbf{x})  \right] \psi^{(3)}(t_{n+1},\mathbf{x}).
\end{eqnarray}
where $T$, the time-ordering operator, has been introduced. The latter orders the argument of the exponential function according to their time argument: from the smaller time, on the right to the larger time, on the left. This is required because $\beta mc^{2}-e\boldsymbol{\alpha} \cdot  \mathbf{A}(t,\mathbf{x})$ does not commute with itself when it is evaluated at different times due to its Dirac structure (the commutator $[h(t),h(t')] \neq 0$ where $h(t) \equiv \beta mc^{2} -e \boldsymbol{\alpha} \cdot  \mathbf{A}(t,\mathbf{x})$). In the following, this $T$-ordering will be omitted ($T \exp (...) \rightarrow \exp(...)$) as this approximation results in an error of $O(\delta t^{3})$ \cite{PhysRevA.59.604}, which has the same (or better) accuracy as the method considered. Therefore, this last step of the splitting is the only one which is not exact: the time-ordering is neglected and the computation of the integration requires a numerical quadrature in general.    

Armed with these analytical solutions, we can now discuss precisely how this system of equation is discretized spatially.

\subsection{Spatial Discretization}

In this numerical method, a finite volume formulation is considered where the space domain is discretized in cubic elements with edges of length $a = \delta x = \delta y = \delta z$. Inside each element, the value of the wave-function is constant ($P_{0}$ type elements). Mathematically, the discretized wave function and electromagnetic field can be written as
\begin{eqnarray}
\label{eq:discr_psi}
 \psi_{h}(t,\mathbf{i}) &=& \sum_{m=1}^{N} \mathbf{1}_{m}(\mathbf{i}) \psi(t,\bar{\mathbf{x}}_{m}) \\
\label{eq:discr_A}
\mathbf{ A}_{h}(t,\mathbf{i}) &=& \sum_{m=1}^{N} \mathbf{1}_{m}(\mathbf{i}) \mathbf{A}(t,\bar{\mathbf{x}}_{m})
\end{eqnarray}
where $N=N_{x}N_{y}N_{z}$ is the total number of elements, $\psi_{h}(t,\mathbf{i})$ and $\mathbf{A}_{h}(t,\mathbf{i})$ are the discretized wave-function and electromagnetic field ($ \mathbf{i} \equiv (i,j,k) \in \mathbb{Z}^{3}$ are indexing the volumes), the function $\mathbf{1}_{m}(\mathbf{i})$ is 1 in volume $m$ indexed by $\mathbf{i}$ and zero outside, while $\bar{\mathbf{x}}_{m}$ is the vector pointing to the center of volume $m$. The latter is defined as 
\begin{eqnarray}
\bar{\mathbf{x}}_{m} = \left(x_{\rm min} + (i+\frac{1}{2})a, y_{\rm min} + (j+\frac{1}{2})a, z_{\rm min} + (k+\frac{1}{2})a\right) 
\end{eqnarray}
where $x_{\rm min},y_{\rm min},z_{\rm min}$ are the lower domain boundary coordinates. 

Eqs. (\ref{eq:discr_psi}) and (\ref{eq:discr_A}) represent the space discretization based on finite volume elements of edge size $a$. On the other hand, Eqs. (\ref{eq:sol_ex_1}) to (\ref{eq:sol_ex_4}) are the time discretization using an operator splitting with a time increment of $\delta t$. To get the full numerical scheme, they need to be combined together. This is performed in the following way.

First, note that \textit{a priori}, the time ($\delta t$) and space ($a$) increments are arbitrary and unrelated to each other. However, on the characteristics curves, the condition $c \delta t = a$ is obeyed (this is the equation of the characteristics where the partial differential equation becomes an ordinary differential equation). This is the reason for the appearance of factors like $(x,y,z) \pm c \delta t$ in the wave function argument of Eqs. (\ref{eq:sol_ex_1}) to (\ref{eq:sol_ex_3}). Choosing specifically this relation between the time and space increments guarantees that the method is unconditionally stable \cite{Lorin2011190}. Moreover, it allows to write $\psi(x\pm c\delta t) \rightarrow \psi_{h}(i\pm K)$ when the system of equation is discretized in space and this is exact, i.e. this discretized scheme reproduces exactly the analytical form in each dimension (up to errors coming from the projection of the grid). This is true for any $K \in \mathbb{N}^{*}$ such that $c \delta t = K a$. In the following, we chose $K=1$ for simplicity.

These facts can be understood in a more physical way. As stated earlier, Eqs. (\ref{eq:sol_ex_1}) to (\ref{eq:sol_ex_3}) are traveling waves moving at velocity $c$. Therefore, during a time $\delta t$, the wave travels a distance $d=c \delta t$. When the wave is discretized, we take its value at the center of each element. By choosing $a=d$, the value of the continuous and discretized wave coincide at every time steps because the information moves from one element center to the neighbor element center. This would not be the case of other choices of space and time increments and this would induce numerical diffusion.

With this choice of space discretization, Eqs. (\ref{eq:sol_ex_1}) to (\ref{eq:sol_ex_4}) become
\begin{eqnarray}
\label{eq:sol_dis_1}
 \psi_h^{n_1}(\mathbf{i}) &=& \frac{1}{2} \biggl\{ [\mathbb{I}_{4} + \alpha_{x}] \psi_h^n(i-1,j,k)   + [\mathbb{I}_{4} - \alpha_{x}] \psi_h^n(i+1,j,k)  \biggr\} \\
\label{eq:sol_dis_2}
 \psi_h^{n_2}(\mathbf{i}) &=& \frac{1}{2} \biggl\{ [\mathbb{I}_{4} + \alpha_{y}] \psi_h^{n_1}(i,j-1,k)   + [\mathbb{I}_{4} - \alpha_{y}] \psi_h^{n_1}(i,j+1,k) \biggr\} \\
\label{eq:sol_dis_3}
 \psi_h^{n_3}(\mathbf{i}) &=& \frac{1}{2} \biggl\{ [\mathbb{I}_{4} + \alpha_{z}] \psi_h^{n_2}(i,k,k-1)  + [\mathbb{I}_{4} - \alpha_{z}] \psi_h^{n_2}(i,j,k+1)  \biggr\}  \\
\label{eq:sol_dis_4}
 \psi^{n+1}_{h}(\mathbf{i}) & =&  \exp \left[ -i \beta mc^{2}\delta t -i \tilde{V}^{n}_{h}(\mathbf{i}) +i \boldsymbol{\alpha} \cdot \tilde{\mathbf{A}}_{h}(\mathbf{i}) \right] \psi^{n_3}_{h}(\mathbf{i})
\end{eqnarray}
where we defined
\begin{eqnarray}
\label{eq:field_dis1}
\tilde{\mathbf{A}}^{n}(\mathbf{x}) &=& e\int_{t_{n}}^{t_{n+1}} d \tau \mathbf{A}(\tau, \mathbf{x}) \\
\label{eq:field_dis2}
\tilde{V}^{n}(\mathbf{x}) &=& e\int_{t_{n}}^{t_{n+1}} d\tau V(\tau,\mathbf{x}) .
\end{eqnarray}
The discretization of $\tilde{\mathbf{A}}$ and $\tilde{V}$ proceeds as in Eq. (\ref{eq:discr_psi}) using the same finite volume formulation. The integrals in the last equations can be evaluated either analytically, if the expression of the electromagnetic field is simple enough, or numerically, using some kind of quadrature. In the latter case, this requires a subdivision of the time increment and thus, is more time consuming. Moreover, if the Dirac solver is coupled to a Maxwell equation solver that computes the time-dependent electromagnetic potential, as in \cite{Lorin2011190}, the evaluation of these integrals requires even more resources. However, if the space-time variations of the electromagnetic field are smaller than the typical space-time variations of the wave-function, that is 
\begin{equation}
 |\nabla \mathbf{A}| \ll  |\nabla \psi| \;\; \mbox{and} \;\;  |\partial_{t} \mathbf{A}| \ll  |\partial_{t} \psi|
\end{equation}
for all $\mathbf{x}$, then we can simplify Eqs. (\ref{eq:field_dis1}) and (\ref{eq:field_dis2}) as
\begin{eqnarray}
\tilde{\mathbf{A}}^{n}(\mathbf{x}) \approx \mathbf{A}^{n}(\mathbf{x}) \delta t \; ; \;
\tilde{V}^{n}(\mathbf{x}) \approx V^{n}(\mathbf{x}) \delta t.
\end{eqnarray}
This approximation was used in \cite{Lorin2011190} and is valid for many cases of interests because the macroscopic electromagnetic field usually involves much larger time-scales as that of the electron. In this work however, we consider only simple electromagnetic fields which can be integrated analytically and thus, the numerical integration is not required. This is also true for certain laser pulse models (see \cite{Bauke2011} for instance for a pulse with a linear turn-on ramp) for which the integral can be found analytically.    

The first exponential function in Eq. (\ref{eq:sol_dis_4}) is the formal expression for a $4\times 4$ matrix. It could be evaluated numerically by performing a similarity transformation that diagonalizes the matrix in the exponential. However, it is more convenient and efficient to use the well-known result for the exponential of Dirac matrices, that is\footnote{This relation can be derived by using the Taylor expansion of the exponential function and the following Dirac matrices properties for any positive integers $n$: $\beta^{2n} = 1, \; \alpha_{i}^{2n} = 1, \; \beta^{2n+1} = \beta, \; \alpha_{i}^{2n+1} = \alpha_{i}$).}
\begin{eqnarray}
 e^{i(\beta F + \boldsymbol{\alpha} \cdot \mathbf{F} )} = \cos(|F|) + i \frac{(\beta F + \boldsymbol{\alpha} \cdot \mathbf{F} )}{|F|} \sin(|F|)
\end{eqnarray}
where $\mathbf{F} = (F_{1},F_{2},F_{3})$, $F,F_{i}$ are arbitrary scalar functions and $|F| \equiv \sqrt{F^{2} + \mathbf{F} \cdot \mathbf{F}}$. From this result, the solution in Eq. (\ref{eq:sol_dis_4}) becomes
\begin{eqnarray}
  \psi^{n+1}_{h}(\mathbf{i}) & =&U(\mathbf{i}) \exp \left[ -i \tilde{V}^{n}_{h}(\mathbf{i}) \right]  \psi^{n_3}_{h}(\mathbf{i}) 
\end{eqnarray}
where $U(\mathbf{i})$ is a matrix given explicitly by
\begin{eqnarray}
 U(\mathbf{i}) \equiv \nonumber \\
\begin{bmatrix}
 \mathrm{c}(A) - i\frac{mc^{2} \delta t}{A} \mathrm{s}(A) & 0 & i \frac{\tilde{A}_{h,z}(\mathbf{i})}{A} \mathrm{s}(A) & \frac{[i\tilde{A}_{h,x}(\mathbf{i}) + \tilde{A}_{h,y}(\mathbf{i})]}{A} \mathrm{s}(A)\\
0 &  \mathrm{c}(A) - i\frac{mc^{2} \delta t}{A} \mathrm{s}(A) &\frac{[i\tilde{A}_{h,x}(\mathbf{i}) - \tilde{A}_{h,y}(\mathbf{i})]}{A} \mathrm{s}(A)& -i \frac{\tilde{A}_{h,z}(\mathbf{i})}{A} \mathrm{s}(A) \\
i \frac{\tilde{A}_{h,z}(\mathbf{i})}{A} \mathrm{s}(A) & \frac{[i\tilde{A}_{h,x}(\mathbf{i}) + \tilde{A}_{h,y}(\mathbf{i})]}{A} \mathrm{s}(A) &  \mathrm{c}(A) + i\frac{mc^{2} \delta t}{A} \mathrm{s}(A) & 0 \\
\frac{[i\tilde{A}_{h,x}(\mathbf{i}) - \tilde{A}_{h,y}(\mathbf{i})]}{A} \mathrm{s}(A)& -i \frac{\tilde{A}_{h,z}(\mathbf{i})}{A} \mathrm{s}(A) & 0 &  \mathrm{c}(A) + i\frac{mc^{2} \delta t}{A} \mathrm{s}(A)
\end{bmatrix}
\end{eqnarray}
In the last expression, we defined
\begin{eqnarray}
 \mathrm{c}(A) \equiv \cos(A) \;\;, \;\; \mathrm{s}(A) \equiv \sin(A) \;\; \mbox{and} \;\; A \equiv \sqrt{(mc^{2} \delta t)^{2} + \tilde{\mathbf{A}}_{h}(\mathbf{i}) \cdot \tilde{\mathbf{A}}_{h}(\mathbf{i})} .
\end{eqnarray}

\subsection{Higher order splitting}

The main numerical error in the scheme presented in the preceding section is due to the operator splitting. As argued in the last section, this scheme is exact in each dimension up to the projection of the wave function and electromagnetic field on the mesh and it can be shown that it is a first order method \cite{Lorin2011190}. This can be improved significantly by choosing higher order splitting schemes like the second order Strang-like splitting scheme given by (we omit the $\mathbf{x}$ in arguments for notational convenience)
\begin{eqnarray}
\begin{array}{lll}
i\partial_{t} \psi^{(1)}(t) = \hat{A} \psi^{(1)}(t), & \psi^{(1)}(t_{n}) = \psi^n,  & t \in [t_{n},t_{n+\frac{1}{4}}) \nonumber \\
i\partial_{t} \psi^{(2)}(t) =\hat{B} \psi^{(2)}(t), & \psi^{(2)}(t_{n}) = \psi^{(1)}(t_{n+\frac{1}{4}}), & t \in [t_{n},t_{n+\frac{1}{2}}) \nonumber \\
i\partial_{t} \psi^{(3)}(t) = \hat{A} \psi^{(3)}(t), & \psi^{(3)}(t_{n+\frac{1}{4}}) = \psi^{(2)}(t_{n+\frac{1}{2}}), & t \in [t_{n+\frac{1}{4}},t_{n+\frac{1}{2}})  \nonumber \\
i\partial_{t} \psi^{(4)}(t) = \hat{C} \psi^{(4)}(t), & \psi^{(4)}(t_{n}) = \psi^{(3)}(t_{n+\frac{1}{2}}), & t \in [t_{n},t_{n+\frac{1}{2}})  \nonumber \\
i\partial_{t} \psi^{(5)}(t) = \hat{D} \psi^{(5)}(t), & \psi^{(5)}(t_{n}) = \psi^{(4)}(t_{n+\frac{1}{2}}),  & t \in [t_{n},t_{n+1}) \nonumber \\
i\partial_{t} \psi^{(6)}(t) =\hat{C} \psi^{(6)}(t), & \psi^{(6)}(t_{n+\frac{1}{2}}) = \psi^{(5)}(t_{n+1}), & t \in [t_{n+\frac{1}{2}},t_{n}) \nonumber \\
i\partial_{t} \psi^{(7)}(t) = \hat{A} \psi^{(7)}(t), & \psi^{(7)}(t_{n+\frac{1}{2}}) = \psi^{(6)}(t_{n+1}), & t \in [t_{n+\frac{1}{2}},t_{n+\frac{3}{4}})  \nonumber \\
i\partial_{t} \psi^{(8)}(t) = \hat{B} \psi^{(8)}(t), & \psi^{(8)}(t_{n+\frac{1}{2}}) = \psi^{(7)}(t_{n+\frac{3}{4}}), & t \in [t_{n+\frac{1}{2}},t_{n+1})  \nonumber \\
i\partial_{t} \psi^{(9)}(t) = \hat{A} \psi^{(9)}(t), & \psi^{(9)}(t_{n+\frac{3}{4}}) =  \psi^{(8)}(t_{n+1}),& t \in [t_{n+\frac{3}{4}},t_{n+1}) \nonumber \\
\end{array} \nonumber \\
\mbox{and} \;\; \psi^{n+1} = \psi^{(9)}(t_{n+1})
\end{eqnarray} 
Here, the time increments are $\delta t /4 = t_{n+\frac{1}{4}} - t_{n}$ for the operator $\hat{A}$, $\delta t/2 = t_{t+\frac{1}{2}}-t_{n}$ for $\hat{B}$ and $\hat{C}$, and finally, $\delta t = t_{n+1} - t_{n} $ for $\hat{D}$.  This kind of splitting induces an error of $O(\delta t^{3})$. Then, the space discretization proceeds as in the lowest order splitting, leading to a second order numerical method \cite{Lorin2011190}. Of course, this can be improved to arbitrary order but this increases the computation time significantly. In this work, we consider only the lowest  order and second order splitting. 

It is also important to note that in 1-D and 2-D, the last equation simplifies considerably because the number of steps in the splitting is reduced. For instance, if one considers the $x-z$ plane in 2-D, it is possible to omit every steps having the operator $\hat{B}$ (the same is true for the lowest order splitting in Eqs. (\ref{eq:split1}) to (\ref{eq:split4})) while the subsequent steps with operator $\hat{A}$ can be combined into one step with a time increment of $\delta t /2$. In 1-D, for the $x$-line, it is possible to omit all steps with operators $\hat{B}$ and $\hat{C}$. In that case, the lowest order and second order splitting are equivalent. Also, the Dirac equation in 1-D reduces to a two-component equation and thus, describes the time-evolution of a bi-spinor. This simplification however is not implemented in the following applications.      

\subsection{Time step size}
\label{sec:timestep}

The time step in the numerical solution of the Dirac equation should be chosen carefully to insure the convergence of the calculation. The limit on the maximum time step size is related to the fermion mass energy $mc^{2}$ because the solution (see Eq. (\ref{eq:sol_ex_4})) includes a term having a form like $\sim e^{imc^{2}t}$. The latter oscillates at a frequency $\sim mc^{2}$, which is very high in comparison to other typical time scales. Therefore, the time step should be chosen such as \cite{PhysRevA.83.063414}
\begin{eqnarray}
\label{eq:time_step}
 \delta t \ll \frac{1}{mc^{2}} \approx 5.0 \times 10^{-5} \; \mbox{a.u.}.
\end{eqnarray}
As will be seen in the examples of Section \ref{sec:validation}, accurate results can be obtained with $\delta t < 1.0 \times 10^{-5}$ a.u..

\section{Fermion-Doubling Problem}
\label{sec:fermi_double}

To our knowledge, the fermion-doubling problem was first encountered in Quantum Field Theory (QFT) on the lattice (see \cite{RevModPhys.55.775,DeGrand:2006rh} for a review of this issue). In these studies, where the goal is the evaluation of path integrals for the evaluation of Euclidian correlators, this problem appears when the fermionic action (from which the Dirac equation is obtained by a variational derivative) is discretized using a symmetric difference scheme for the space derivatives\footnote{We define the symmetric difference scheme as $\partial_{x}u(x) \rightarrow \frac{u(x+a)-u(x-a)}{2a}$, where $a$ is the lattice spacing.}. The latter choice is required to keep the hermiticity of the action and of the momentum operator. This however modifies the dispersion relation, which becomes (in 1D and assuming a free and massless theory)\cite{PhysRevD.26.468}
\begin{equation}
 E^{2} \sim \frac{1}{a^{2}} \sin^{2} \left( pa\right).
\end{equation}
where $a$ is the lattice spacing, $E$ is the fermion energy and $p$ is the momentum. It is then easy to show that the resulting discrete fermionic propagators have two poles in the Brillouin zone at $p=0$ and $p=\frac{\pi}{a}$, representing two different fermionic states (in $D$ dimensions, we have $2^{D}$ fermionic states \cite{PhysRevD.26.468}). Therefore, the discretized action leads to a theory describing many fermions while its continuum counterpart has only one fermionic species. Finding a consistent way to resolve this discrepancy between the continuum and discrete theories is still an open problem in Lattice QFT. Under fairly general assumption, it was proven in the Nielsen-Ninomiya theorem \cite{Nielsen198120} that it is actually impossible to keep both the hermiticity of the discretized fermionic action and the chiral symmetry without adding the fermion doublers.  




This discrepancy between the continuum and the discretized versions of the action also appears at the level of the equation of motion (for coordinate space discretizations), as discussed thoroughly in the literature \cite{RevModPhys.55.775,PhysRevD.26.468,DeGrand:2006rh}. From the QFT lattice studies, two methods have emerged which are free from spurious states: the Wilson and the Staggered fermions equation of motion. Note however that both breaks chiral symmetry \cite{RevModPhys.55.775}. In the Wilson discretization, a new term is added which cancels the second minima of the dispersion relation. It is interesting to note that this term is actually the same as the artificial viscosity term appearing in advection equation discretization schemes (such as in Lax-Friedrichs scheme), which is well-known in numerical analysis. In the Staggered approach, the fermion doubling is avoided by changing the size of the Brillouin zone. This is performed by letting the wave function components belong to different lattice points \cite{RevModPhys.55.775}. In general, real space discretization schemes of the time-dependent Dirac equation used for applications are variations of these two methods (see \cite{PhysRevA.40.5548,PhysRevA.79.043418}), thus circumventing the fermion doubling problem. Otherwise, unphysical behavior may be seen in the numerical solution. For instance, in a finite element discretization of the Dirac equation where the doubling appears, it was shown in \cite{Muller1998245} that an electron wave packet could move at a velocity greater than the speed of light. This stresses the importance of this issue. 

We now analyze our numerical method and claim that it does not suffer from this fermion-doubling problem.

\subsection{1-D}

We will now show that our numerical method is free from these spurious states. For simplicity, let us first consider the 1-D massless Dirac equation without external field (the 3-D case will be treated in the next subsection), given by
\begin{eqnarray}
i \partial_{t} \psi(t,x) = - ic \alpha_{x} \partial_{x} \psi(t,x) , 
\end{eqnarray}
which actually corresponds to the first step of our operator splitting method, i.e. Eq. (\ref{eq:split1}). The solution to this equation can be computed by taking the Fourier transform in space. From this procedure, we get
\begin{equation}
\label{eq:sol_1D_doub}
 \widehat{\psi}(t+\delta t, p_{x}) = e^{-i c \alpha_{x} p_{x} \delta t } \widehat{\psi}(t, p_{x})
\end{equation}
where the widehat denotes the Fourier-transformed function and $p_{x}$ is the momentum in $x$.

Using our discretization method described in the last section, we have that
\begin{eqnarray}
 \psi^{n+1}_{h}(i) &=& \frac{1}{2} \left\{ [\mathbb{I}_{4} + \alpha_{x}] \psi^{n}_{h}(i-1)  + [\mathbb{I}_{4} - \alpha_{x}] \psi^{n}_{h}(i+1)  \right\}. 
\end{eqnarray}
We then take the discrete Fourier transform in space to get
\begin{eqnarray}
 \widehat{\psi}^{n+1}_{h}(k) &=& \frac{1}{2} \left\{ [\mathbb{I}_{4} + \alpha_{x}] e^{-\frac{2\pi i}{N_{x}} k}  + [\mathbb{I}_{4} - \alpha_{x}] e^{ \frac{2\pi i}{N_{x}} k} \right\}\widehat{\psi}^{n}_{h}(k)  \nonumber \\
\label{eq:sol_dis}
&=& e^{-i \alpha_{x} \frac{2 \pi k}{N_{x}}}\widehat{\psi}^{n}_{h}(k) = e^{-i c \alpha_{x} p_{k} \delta t}\widehat{\psi}^{n}_{h}(k) 
\end{eqnarray}
where $N_{x}$ is the number of elements, $p_{k}$ is a discrete momentum  with index $k \in [0,N_{x}-1]$ and the widehat now denotes the discrete Fourier-transformed function. To obtain the last equality, we used the fact that in our discretization, we have $\delta x = c \delta t$. This clearly has the same form as the analytical solution Eq. (\ref{eq:sol_1D_doub}) except that the momentum $p_{k} \equiv \frac{2\pi k}{N_{x} \delta x}$ now takes discrete values. Also, we note that by comparing Eqs. (\ref{eq:sol_dis}) and (\ref{eq:sol_1D_doub}), the exponential function is just the evolution operator. Therefore, the discrete Hamiltonian is $H_{k} = c \alpha_{x} p_{k}$ from which the energy spectrum can be computed by finding its eigenvalues \cite{PhysRevD.26.468}. This yields a dispersion relation given by
\begin{equation}
 E_{k} = \pm c p_{k} 
\end{equation}
where $E_{k}$ is the discrete energy. This is very similar to the continuum case except that the momentum and energy take discrete values. The relationship between them is linear as in the continuum and therefore, there is no fermion-doubling. This result is actually not surprising as our technique is based on the exact solution of the Dirac equation.

\subsection{3-D}

The 3-D case is very similar to the 1-D case. We start by computing the analytical solution in the continuum (without mass and without external field) by using the Fourier transform method. We get
\begin{eqnarray}
\label{eq:sol_3D_doub}
 \widehat{\psi}(t+\delta t, \mathbf{p}) &=& e^{-i c (\alpha_{x} p_{x} + \alpha_{z} p_{z} + \alpha_{z} p_{z}) \delta t } \widehat{\psi}(t, \mathbf{p})  \nonumber \\ 
&=& \left[ \mathbb{I}_{4} -i\hat{H}_{0}\delta t \right]\widehat{\psi}(t, \mathbf{p})   + O[(\delta t)^{2}].
\end{eqnarray}
where $\hat{H}_{0}$ is the massless free Hamiltonian (with $m = V = \mathbf{A} = 0$). For the discrete part, each coordinate is treated sequentially as in Eq. (\ref{eq:split1}) to (\ref{eq:split3}) because we are using an operator splitting method\footnote{This analysis is for the first order splitting. The analysis can be generalized to higher order splitting in the same way.}. Therefore, we get
\begin{eqnarray}
 \widehat{\psi}^{n+1}_{h}(\mathbf{k}) = e^{-i c \alpha_{z} p_{k_{z}} \delta t}e^{-i c \alpha_{y} p_{k_{y}} \delta t}e^{-i c \alpha_{x} p_{k_{x}} \delta t}\widehat{\psi}^{n}_{h}(\mathbf{k}) .
\end{eqnarray}
This of course cannot be compared directly to Eq. (\ref{eq:sol_3D_doub}) because of the anti-commuting nature of Dirac matrices. However, if we also expand the exponential, we get an equation similar to the continuum case at the order $O(\delta t)$, allowing us to determine the discrete Hamiltonian and energy spectrum by inspection: 
\begin{eqnarray}
 H_{\mathbf{k}} = c \alpha_{z} p_{k_{z}} + c \alpha_{y} p_{k_{y}} + c \alpha_{x} p_{k_{x}} + O(\delta t) ,\\
E_{\mathbf{k}} = \pm c \sqrt{p_{k_{x}}^{2} + p_{k_{y}}^{2} + p_{k_{z}}^{2}} + O(\delta t) .
\end{eqnarray}
This is again a discrete dispersion relation without fermion-doubling problem and this proves our assertion that our numerical method is free from these spurious states. Note however that in the 3-D case, the dispersion relation is not exact as in 1-D due to the operator splitting, which necessitates the addition of $O(\delta t)$ terms and higher. The order of the numerical error on the dispersion relation will obviously vary with the splitting order. Nevertheless, this will not change our conclusion concerning fermion-doubling. For instance, in the first order splitting, the $O(\delta t)$ terms in the Hamiltonian are $H_{\mathbf{k}}^{O(\delta t)} = c^{2}p_{k_{x}}p_{k_{y}}[\alpha_{y},\alpha_{x}] + c^{2}p_{k_{x}}p_{k_{z}}[\alpha_{z},\alpha_{x}] + c^{2}p_{k_{y}}p_{k_{z}}[\alpha_{z},\alpha_{y}]$ which are linear functions of momenta and thus, do not induce fermion doubling.  


\section{Implementation and performance}
\label{sec:perf}

A computer program using the method described in Section \ref{sec:num_meth} to solve the time-dependent Dirac equation was written in C++. Given its relative simplicity, it is possible to implement the 1-D, 2-D and 3-D cases easily; changing the number of dimensions requires only minor modifications to the code, amounting to the adjustment of some array sizes and to the addition (or deletion) of the action of certain operators defined in Eqs. (\ref{eq:op_A}) to (\ref{eq:op_C}). The first order and second order splitting are included; higher order schemes could also be easily added but require more computation time and are not implemented at the moment. The number of operations per time iteration $n_{\rm op}$ scales like $n_{\rm op} \sim O(N)$ where $N$ is the number of elements. This is slightly better than spectral methods in which the computation of the Fast Fourier Transform scale like $n_{\rm op} \sim O(N\ln N)$. In the latter however, there are less levels in the splitting scheme because the action of the operator $\boldsymbol{\alpha} \cdot \mathbf{p}$ is computed in one step instead of three for our numerical method (for lowest order splitting). This results in a potentially smaller scaling prefactor.  Nevertheless, the strength of our method is that it can be easily parallelized using a domain decomposition strategy, which leads to a very scalable and efficient computation tool (see Section \ref{subsec:para} for the parallel efficiency evaluation).    

%
%
%


\subsection{Parallelization}
\label{subsec:para}

The algorithm is parallelized by using a domain decomposition strategy: the whole domain is divided in subdomains and the mesh data in each of these subdomains is sent to a different process. The communication between these subdomains, which is required in the computation by the elements close to the subdomain boundaries, is performed using the Message Passing Interface (MPI). 

\begin{figure}
\centering
\includegraphics[width=0.9\textwidth]{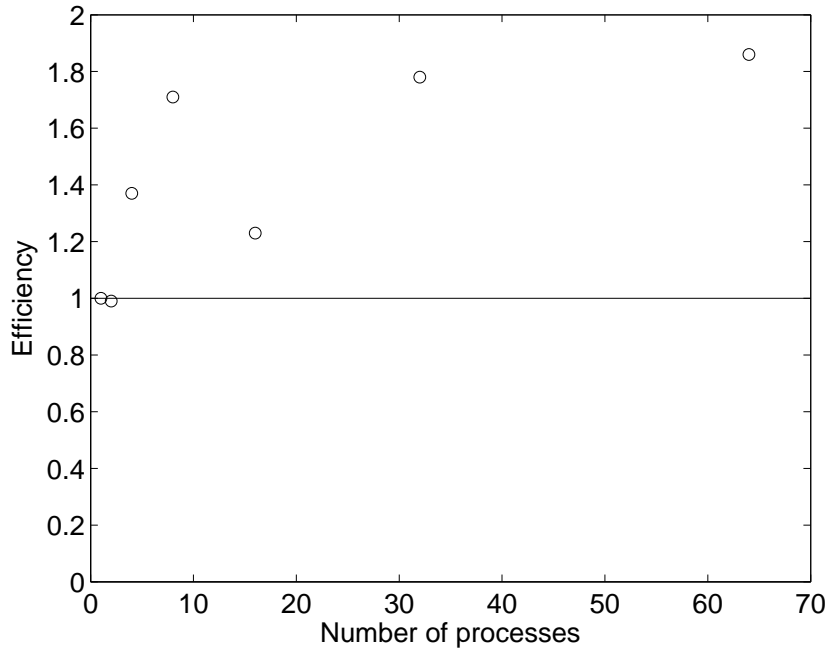}
\caption{Parallelization efficiency as a function of the number of processes.}
\label{fig:par_perf}
\end{figure}

To evaluate the efficiency of our parallelization procedure, we look at the time-evolution of a 3-D Gaussian wave packet. A mesh of $256 \times 256 \times 256$ elements is chosen and the wave function is evolved in time for 500 time iterations (this is the larger problem that can be runned serially using our computer resources). The number of processes is varied while keeping the computation parameters fixed, thus evaluating the strong scaling. The computation time is recorded for each process number and used to evaluate the computation efficiency given by $e = \frac{T_{1}}{pT_{p}}$ where $T_{p}$ is the computation time for $p$ processes. This value, usually between 0 and 1, characterizes the quantity of work done by each processes for the actual computation in comparison to communication times. For an ideal system with linear speedup, we should have $e \approx 1$. All the computations are performed on the MAMMOUTH cluster\footnote{The MAMMOUTH is a cluster of Intel Xeon 64 bit, 3.6 GHz CPUs with an Infiniband network SDR Cisco-Topsprin non-blocking having a bandwidth of 800 MB/sec.}. The results for the scaling properties are shown in Fig. \ref{fig:par_perf}.

In this figure, it is shown that our numerical method have a super linear speedup, i.e. $e > 1$, for the problem considered (3-D Gaussian wave packet) and on the computer used. This at first may seem unexpected: one would expect the parallel version efficiency to be smaller than 1 because at each time step, processes spend time to communicate with their neighbors, which does not happen in serial calculation. The most likely explanation for this phenomenon is the cache effect \cite{wiki:xxx}: as the number of processes is increased, so is the cache memory (the latter is much faster than ordinary memory). Thus, a larger percentage of the problem can be fitted in this memory and this decreases the memory access time. The overall effect is to reduce the computation time significantly. This however implies that the efficiency should diminish for larger mesh where this cache effect becomes less important. This was observed in calculations using larger mesh where doubling the number of processes (for instance, from 64 to 128) resulted approximately in halving the computation time (in these cases, a serial computation was not possible). Therefore, the efficiency in that case is much closer to 1.

\section{Applications to simple systems}
\label{sec:validation}

In this section, some simple systems are studied to validate the numerical method and to show its strength. Many simple physical systems are analyzed which allow a comparison between analytical formulas and numerical results. These are the Gaussian wave packet evolution and the Klein paradox. 

All the computations are performed in atomic units where the electron mass is given by $m=1.0$ and the speed of light is $c = \alpha^{-1}$ where $\alpha$ is the fine structure constant given by $\alpha \approx 1/137.0359895$. Note that in these units, we also have the electron charge $|e|=1$ and the Plank constant $\hbar = 1.0$.  

In all the examples considered in the following, the wave function at the boundaries $\partial \Omega$ is set to $\left. \psi(\mathbf{x},t)\right|_{\partial \Omega} = 0$, which corresponds to a Dirichlet boundary condition. This of course can induce spurious reflections when the wave function reaches the boundary, and this was actually verified empirically. However, by choosing a large enough domain, this effect is reduced considerably and these reflections can be neglected. We are presently working on the implementation of more sophisticated boundary conditions (transparent or absorbing).

\subsection{Time Evolution of Gaussian Wave Packet}

Gaussian wave packets in vacuum are among the most simple systems that can be studied using a wave equation. For this reason, it has been studied extensively in many contexts, whether as a check for the consistency of numerical methods \cite{PhysRevA.59.604,Mocken2008868,Mocken2004558}, to compare with non-relativistic evolution \cite{Su:98} or to study the characteristic features of their time-evolution \cite{PhysRevA.63.032107,2004quant.ph..9079T,PhysRevA.82.052115}. Physically, they represent the localization of a free electron and thus, are very important in many applications. In this part of this work, we are analyzing the time evolution of massless and massive wave packets in 1-D and 2-D to verify and validate our numerical method and its implementation.

\subsubsection{Massless Wave Packet in 2-D}

In this section, the time evolution of a free massless Gaussian wave packet is considered. The motivation for studying such a simple systems is twofold. First, it is possible to derive analytical solutions (at certain specific space-time points) for the time-dependent wave-function in this case, allowing to validate and study the numerical method previously developed. Second, because it does not have a mass term nor electromagnetic potential, it allows us to test the first steps of the splitting and to use a larger grid (the condition in Eq. (\ref{eq:time_step}) does not have to be fulfilled).

The starting point of this analysis is the following initial wave packet equation in 2-D \cite{Itzykson:1980rh} 
\begin{equation}
\label{eq:psi_init_2DWP}
 \psi(t=0, x, z) = \mathcal{N} \begin{bmatrix} 1 \\ 0 \\ 0 \\ 0 \end{bmatrix} e^{- \frac{x^{2} + z^{2}}{4 (\Delta)^{2}}}
\end{equation}
where $\mathcal{N}$ is a normalization constant and $\Delta$ characterizes the Gaussian width. This wave packet represents a spin-up massless electron. The analytical solution of the free massless Dirac equation with (\ref{eq:psi_init_2DWP}) as an initial condition can be found for two specific space-time points:
\begin{enumerate}
 \item When $x=z=0$, we have:
\begin{eqnarray}
 \psi_{1}(t,0,0) &=& \mathcal{N} \left [1 -    \sqrt{\pi} \frac{t}{2 \Delta} e^{-\frac{t^{2}}{4\Delta^2}} \mathrm{erfi}\left( \frac{t}{2 \Delta }\right) \right] \\
 \psi_{2}(t,0,0) &=& \psi_{3}(t,0,0) = \psi_{4}(t,0,0) = 0 
\end{eqnarray}
where $\mathrm{erfi}(z)$ is the imaginary error function\footnote{The imaginary error function is defined as $\mathrm{erfi}(z) = -\frac{2i}{\sqrt{\pi}} \int_{0}^{iz}e^{-t^{2}}dt$. It is related to the error function as $\mathrm{erfi}(z) = -i\mathrm{erf}(iz)$.}.
\item When $ r \equiv \sqrt{x^{2} + z^{2}} = ct$, we have:
\begin{eqnarray}
 \psi_{1}(t,r = ct) &=&  \mathcal{N} \; _{2}F_{2}\left( \frac{1}{4},\frac{3}{4} ; \frac{1}{2}, \frac{1}{2} ; -\frac{t^2}{\Delta^2} \right) \\
 \psi_{2}(t,r = ct) &=& 0 \\
 \psi_{3}(t,r = ct) &=& -\mathcal{N} t^{2} \sin(\phi) \; _{2}F_{2}\left( \frac{5}{4},\frac{7}{4} ; \frac{3}{2}, \frac{5}{2} ; -\frac{t^2}{\Delta^2} \right) \\
 \psi_{4}(t,r = ct) &=& -\mathcal{N} t^{2} \cos(\phi) \; _{2}F_{2}\left( \frac{5}{4},\frac{7}{4} ; \frac{3}{2}, \frac{5}{2} ; -\frac{t^2}{\Delta^2} \right) 
\end{eqnarray}
where $_{2}F_{2}$ is the generalized hypergeometric series\footnote{The generalized hypergeometric series is defined as $_{p}F_{q}(a_{1},...,a_{p};b_{1},...,b_{q};x) = \sum_{k=0}^{\infty} \frac{(a_{1})_{k} ... (a_{p})_{k}}{(b_{1})_{k} ... (b_{q})_{k}} \frac{x^{k}}{k!}$ with the Pochhammer symbol defined as $(a)_{k} = a(a+1)...(a+k-1)$.} and $\phi = \arctan (z/x)$ is the polar angle in real space.
\end{enumerate}
The calculational details necessary to obtain these solutions are relegated to \ref{app:sol_2D_WP}. 

\begin{figure}[h]
\centering
\includegraphics[width=0.90\textwidth]{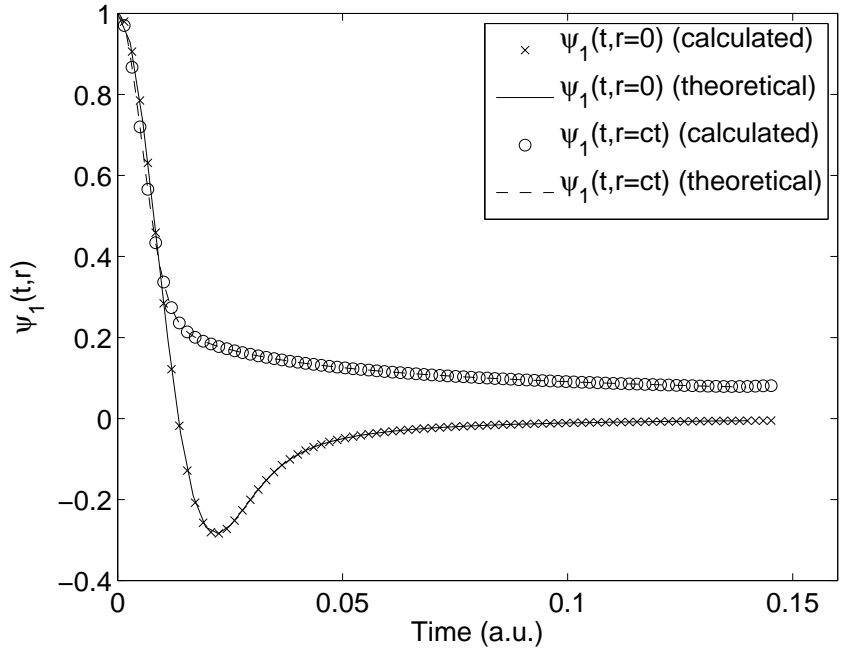}
\caption{Time evolution of the first wave function component for a massless wave packet.}
\label{fig:time_evo_2DWP}
\end{figure}

The results are shown in Fig. \ref{fig:time_evo_2DWP} for the two points considered and for the first wave function component. The domain used in the calculation had a size of $40 \times 40$ a.u., and was discretized in 1000 $\times$ 1000 elements using an order 1 splitting. The theoretical and calculated results show a very good agreement.  

\subsubsection{Wave Packet in 1-D}

In this section, the time evolution of a massive wave packet is analyzed. The calculation is similar to the one performed in the last section, except that the mass term is now considered and the 1-D case is calculated. As discussed in Section \ref{sec:timestep}, this requires a smaller time step size and consequently, a smaller element size (remember that in our numerical method, we have $c \delta t = a$). The expression of the initial wave-packet is given by Eq. (\ref{eq:psi_init_2DWP}) but letting $z=0$ ($\psi_{\rm 1D}(t=0,x) = \psi_{\rm 2D}(t=0,x,z=0)$). Again, it represents a spin-up massive free electron. 

An analytical solution for the wave packet time evolution is computed in \ref{app:sol_1D_WP} and is compared with the result obtained from our numerical method. Integrals in Eqs. (\ref{eq:WP_an_1d_1}) and (\ref{eq:WP_an_1d_4}) are calculated in Matlab using a very high order integration scheme (adaptive quadrature scheme based on a Gauss-Kronrod pair (15th and 7th order formulas)).

\begin{figure}[h]
\centering
\includegraphics[width=0.90\textwidth]{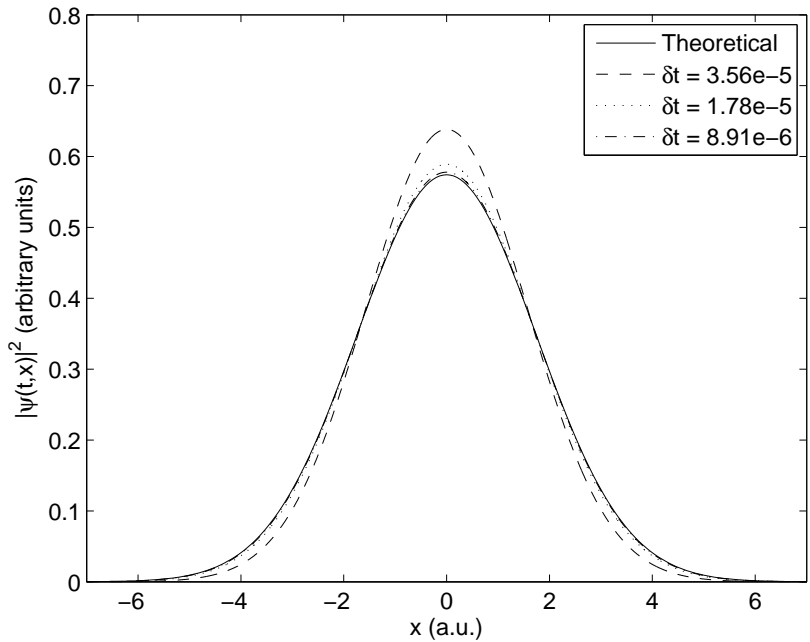}
\caption{Massive 1-D Gaussian wave packet at $t = 2.85$ a.u. for different time step sizes. The theoretical curve is obtained by integrating numerically Eqs. (\ref{eq:WP_an_1d_1}) and (\ref{eq:WP_an_1d_4}). Note that the theoretical result and the third time step ($\delta t_{3} = 8.91 \times 10^{-6}$) are overlapping and thus, are not distinguishable. }
\label{fig:gauss_1d}
\end{figure}

The calculation using our numerical method is performed using the first order splitting and for different element/time step size. The domain boundaries are set to $\pm 10$ a.u., the wave packet initial position is $x_{0} = 0.0$ a.u. and its width is fixed to $\Delta = $ 1.0 a.u.. We consider three time step size: $\delta t_{1} = 3.56 \times 10^{-5}$ a.u., $\delta t_{2} = 1.78 \times 10^{-5}$ a.u. and $\delta t_{3} = 8.91 \times 10^{-6}$ a.u., corresponding to 4096, 8192 and 16284 elements respectively. The results at time $t = 2.85$ a.u. are shown in Fig. \ref{fig:gauss_1d}. This figure shows how critical the choice of time step is, as discussed previously. However, if the condition in Eq. (\ref{eq:time_step}) is fulfilled, as in the case $\delta t_{3}$, the analytical and numerical results are in very good agreement. 

\subsubsection{Wave Packet in 2-D}

\begin{figure}[h]
\centering
\includegraphics[width=0.90\textwidth]{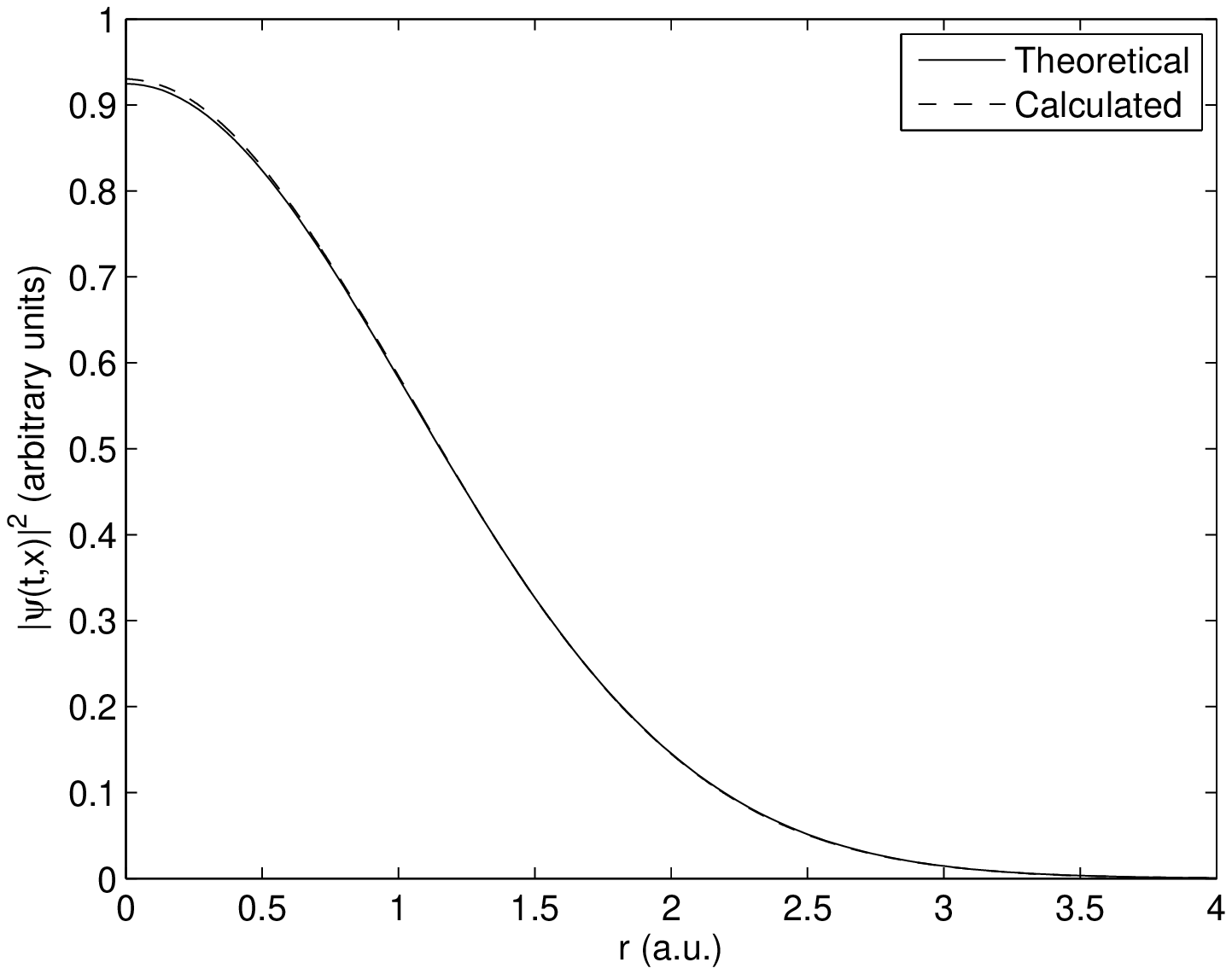}
\caption{Massive 2-D Gaussian wave packet at $t = 0.57$ a.u.. Here, the polar coordinate $r = \sqrt{x^{2}+z^{2}}$ is used because the probability density is symmetric in the azimuthal angle. The results shown in the figure corresponds to the positive $x$-axis ($z=0$ and $x>0$). The calculated results overlap with the theoretical one. }
\label{fig:gauss_2d}
\end{figure}

\begin{figure}[h]
\centering
\includegraphics[width=0.90\textwidth]{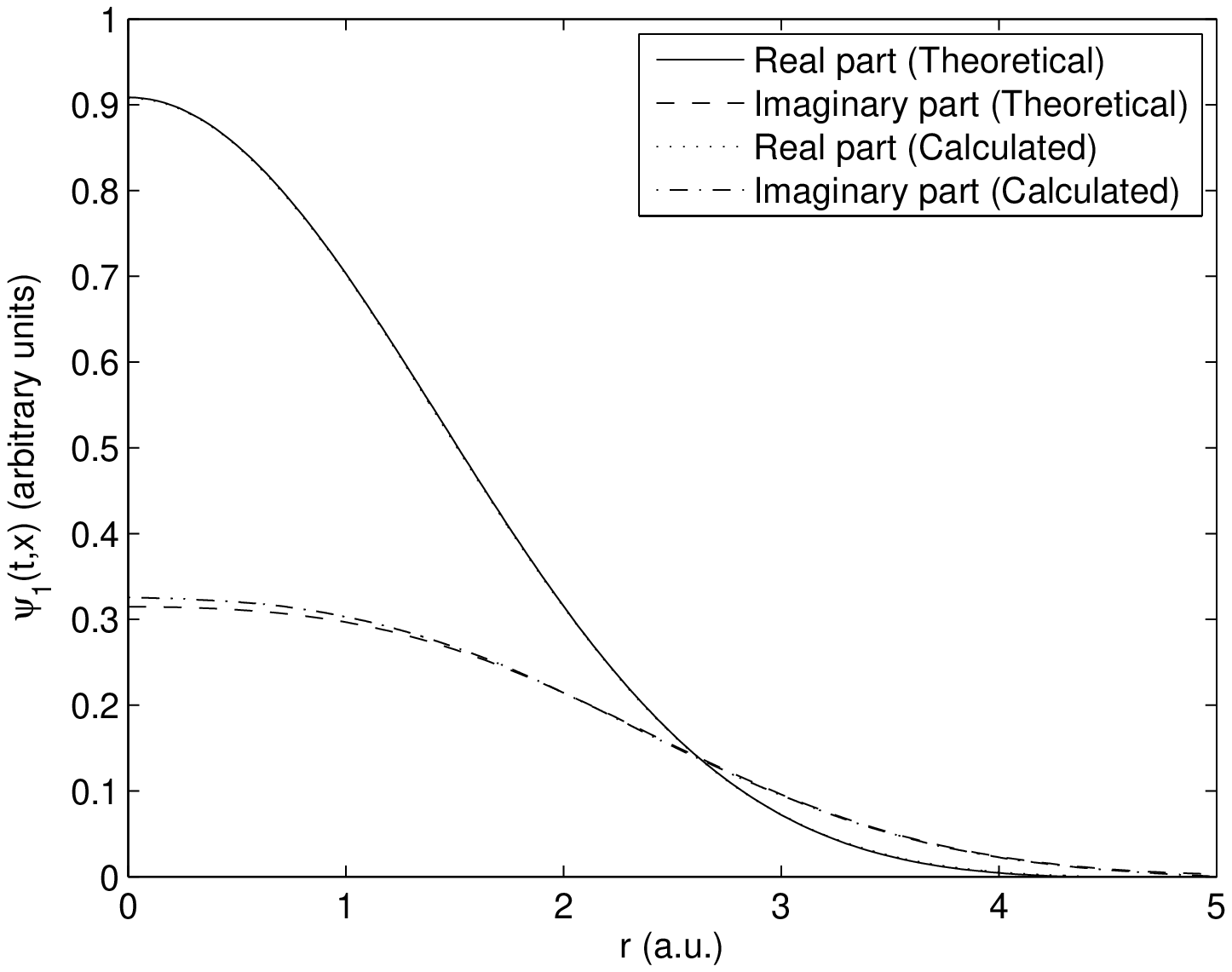}
\caption{Massive 2-D Gaussian wave packet at $t = 0.57$ a.u.. Here, the polar coordinate $r = \sqrt{x^{2}+z^{2}}$ is used because $\psi_{1}$ is symmetric in the azimuthal angle. The results shown in the figure corresponds to the positive $x$-axis ($z=0$ and $x>0$). The calculated results overlap with the theoretical one.}
\label{fig:psi1_2d}
\end{figure}

The same calculation as in the last section is carried in 2-D where the initial wave function is given by Eq. (\ref{eq:psi_init_2DWP}). The analytical solution is calculated in \ref{app:sol_2D_WP} and the integrals in Eqs. (\ref{eq:WP_an_2d_1}), (\ref{eq:WP_an_2d_3}) and (\ref{eq:WP_an_2d_4}) are again calculated in Matlab using the same numerical method. The domain investigated is subdivided into $8192 \times 8192$ elements and its boundaries are positioned at $\pm 5$ a.u. (in both $x$ and $z$). The second order scheme is used, making for a time step of $\delta t = 1.78 \times 10^{-5}$ a.u.. The wave packet initial position is $x_{0} = 0.0$ and its width is given by $\Delta = 1.0$ a.u.. The wave function is evolved for 32000 time iterations to a final time of $t = 0.57$ a.u. and the results are shown in Fig. \ref{fig:gauss_2d}. The theoretical and calculated results are in good agreement for the wave function density $\rho(x,t)$. This is also true when considering the wave function spinorial components as shown in Fig. \ref{fig:psi1_2d} (for the first component only).

\subsubsection{Travelling Wave Packet in 1-D}

So far, the calculations performed considered only wave packets at rest. In this section, we consider the case where the wave packet is given an initial momentum, allowing it to travel on the $x$-axis. The main reason for studying this system is to investigate the phase of the wave function and to determine if the latter is reproduced accurately by the numerical method. Initially, the wavefunction is chosen to be a Gaussian wave packet as
\begin{equation}
\label{eq:trav_wave}
 \psi(t=0, x) = \mathcal{N} \begin{bmatrix} 1 \\ 0 \\ 0 \\ C \end{bmatrix} e^{ik_0 x} e^{- \frac{(x-x_{0})^{2}}{4 (\Delta )^{2}}} \;\; \mbox{with} \;\; C \equiv \frac{c k_{0}}{mc^2 + \sqrt{m^2c^4 + c^{2}k_{0}^2}}
\end{equation}
where $k_{0}$ is the wave packet momentum, $\Delta$ is the wave packet spreading and $x_{0}$ is its initial position. This corresponds to a superposition of positive and negative energy free solutions propagating on the $x$-axis.

The analytical solution for the wave packet time evolution is computed in \ref{app:sol_1D_WP} and is compared with the result obtained from our numerical method. We consider only the first wave function component given by Eq. (\ref{eq:WP_an_1d_1_t}). The integral in this equation is calculated in Maple using a numerical scheme based on an adaptive Gauss-Kronrod quadrature (with Gauss 30-point and Kronrod 61-point rules), well suited for oscillatory integrands.

\begin{figure}[h]
\centering
\includegraphics[width=0.90\textwidth]{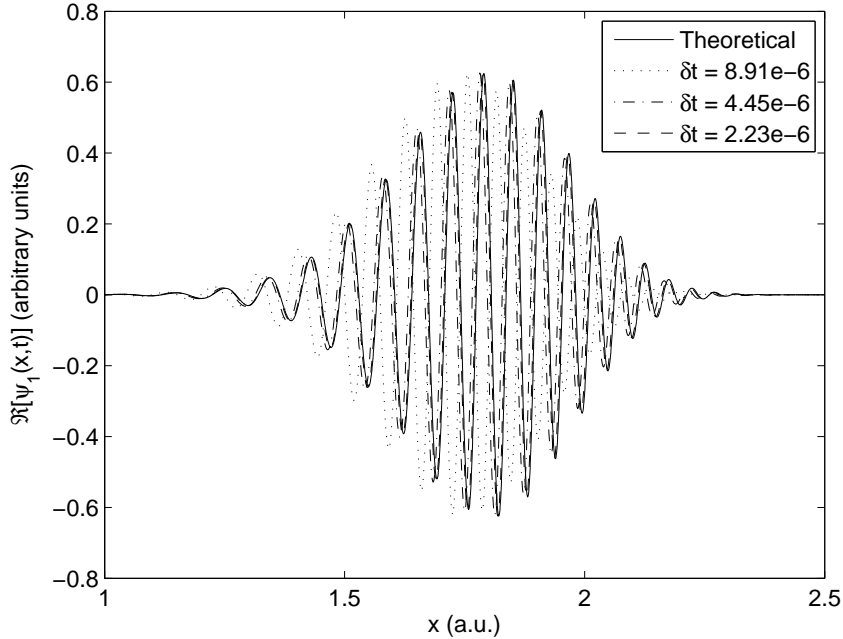}
\caption{Real part of $\psi_{1}$ for a massive 1-D traveling Gaussian wave packet at $t = 0.022$ a.u.. The theoretical curve is obtained by integrating numerically Eq. (\ref{eq:WP_an_1d_1_t}). Note that the theoretical result and the third time step ($\delta t_{3} =  2.23 \times 10^{-6}$) are overlapping and thus, are not distinguishable. }
\label{fig:psi1_1d_trav}
\end{figure}

The calculation using our numerical method is performed using the first order splitting and for different element/time step size. The domain boundaries are set to $\pm 5$ a.u., the wave packet initial position is $x_{0} = 0.0$ a.u.,  its width is fixed to $\Delta = $ 0.05 a.u. and its initial momentum to $k_{0} = $ 100 a.u.. We consider three time step sizes: $\delta t_{1} = 8.91 \times 10^{-6}$ a.u., $\delta t_{2} = 4.45 \times 10^{-6}$ a.u. and $\delta t_{3} = 2.23 \times 10^{-6}$ a.u., corresponding to 4096, 8192 and 16284 elements respectively. The results for $\Re \psi_{1}$ at time $t = 0.022$ a.u. are shown in Fig. \ref{fig:psi1_1d_trav} ($\Im \psi_{1}$ and the other components have similar behavior). This figure shows that the wave function obtained from the numerical method is not as accurate as in the previously studied cases: for the same grid size, the phase error is more important. Thus, a smaller time step is required to reproduce the theoretical wave function precisely ($\sim$4 times smaller). Nevertheless, the analytical and numerical results are in very good agreement for $\delta t = \delta t_{3}$, which confirms the convergence of the numerical scheme. 

\subsection{Klein Paradox}

The ``Klein paradox'' concerns the reflection and transmission of plane wave solution on a potential step of height $V_{0}$ \cite{springerlink:10.1007/BF01339716}. In the non-relativistic (Schr\"odinger) case, when the energy of the incident wave is lower than the potential step, the solution in region where $V(x)$ is non-zero (in $\mathbb{R}^{+}$) is a decaying exponential. As a consequence, most of the wave function is reflected, resulting in a very small transmission coefficient. In the relativistic (Dirac) case, there is a regime, when $E < V_{0} - mc^{2}$, where a new phenomenon appears: a plane wave solution is now possible, yielding a non-negligible transmission coefficient even if $E<V_{0}$. This ``paradox'' has been interpreted in the context of the Dirac see picture \cite{PhysRevLett.92.040406,Greiner:1987,Greiner:1985} and in second quantization \cite{PhysRevLett.92.040406,Greiner:1985} as the production of antimatter on the potential boundary (see also \cite{Dombey199941,Katsnelson2006} for recent views on the subject and possible experimental verification). Thus, the transmitted part is related to the negative energy solution which describes anti-fermions, while the incoming and reflected parts are the electron wave function. In this section, we consider the scattering on a step potential of travelling wave packets in 1-D and 2-D. 

\subsubsection{1-D}

In this subsection, the Klein paradox is analyzed along the same lines as in \cite{PhysRevA.59.604,PhysRevLett.92.040406}, but using the numerical method developed in the preceding sections. This numerical test is chosen because it is one of the few existing analytical solutions of the Dirac equation for time-dependent systems. The calculation is performed in 1-D here and in 2-D in the next section.

Initially, the wavefunction is chosen to be the traveling Gaussian wave packet given in Eq. (\ref{eq:trav_wave}). Also, a smooth potential barrier is considered to avoid numerical problems related to discontinuous functions. The potential is given by
\begin{equation}
 V(x) = \frac{V_{0}}{2} \left[ 1 + \tanh \left( \frac{x}{L} \right)  \right]
\end{equation}
where $V_{0}$ is the magnitude of the potential and $L$ characterizes the gradient and width of the step. This potential has already been studied for incident plane wave solutions and one finds a transmission coefficient given by \cite{PhysRevLett.92.040406,springerlink:10.1007/BF01339461}
\begin{equation}
\label{eq:trans_co}
 \mathcal{T} =- \frac{\sinh(\pi k L) \sinh(\pi k' L) }{\sinh \left[ \pi \left(\frac{V_0}{c} + k + k' \right) \frac{L}{2} \right] \sinh \left[ \pi \left(\frac{V_0}{c} - k - k' \right) \frac{L}{2} \right]}
\end{equation}
where $k = \frac{1}{c} \sqrt{(E_k - V_0)^2 - m^2 c^4}$, $k' = - \frac{1}{c} \sqrt{E_k^2 - m^2 c^4}$ and $E_{k} = \sqrt{k_0^{2} c^2 + m^2 c^{4}}$. This transmission coefficient is valid, strictly speaking, for monochromatic plane wave solutions. However, the latter are not very convenient for numerical calculations because they necessitate the addition of sources (and absorbers) on the numerical domain boundaries. By choosing $\Delta^{-1} \ll k_{0} $ (a Gaussian peaked on momentum $k_{0}$), the error due to the utilization of a Gaussian wave packet instead of a monochromatic wave function is negligible. 

The parameters for the initial wave packet and step potential are the same as in \cite{PhysRevA.59.604}: the momentum and width of the wave packet are $k_{0} = 106.$ a.u. and $\Delta  = 1.0$ a.u. respectively while the width of the step is $L = 10^{-4}$. The simulation domain is $-20 \; \mbox{a.u.} < x < 20$ a.u. which allows us to place the initial wave packet at an initial position $x_{0} = -10$ a.u.. For the first order splitting, the domain is divided into 65536 elements such that each element have a width of $a=6.10 \times 10^{-4}$ (and $\delta t = 4.45 \time 10^{-6}$ a.u.) while for second order splitting, we have twice as much elements (131072 elements with $a=3.05 \times 10^{-4}$ and the same time increment). 

\begin{figure}[h]
\centering
\includegraphics[width=0.7\textwidth]{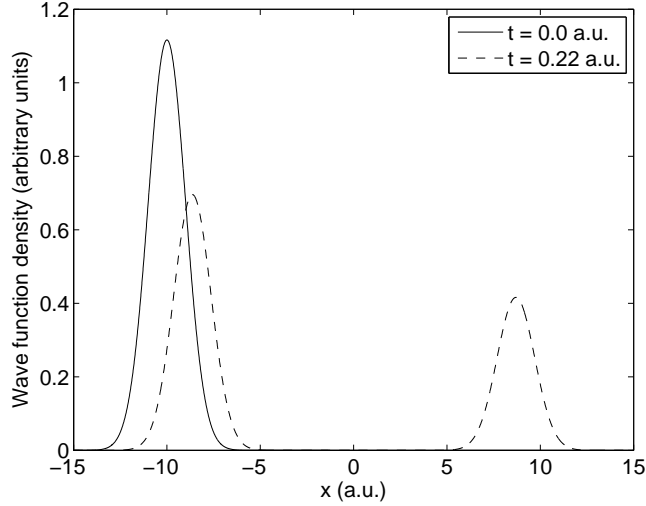}
\caption{Typical wave density ($| \psi(x)|^{2}$) before and after the interaction of the wave packet with the potential barrier. The latter is situated at $x=0.0$ a.u.. The transmitted part is in the region $x>0$, while the initial and reflected wave functions are in $x<0$. }
\label{fig:klein_density}
\end{figure}

\begin{figure}
\centering
\includegraphics[width=0.70\textwidth]{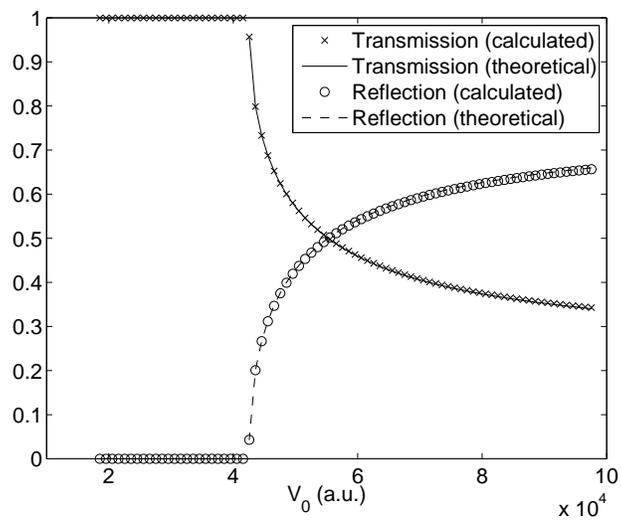}
\caption{Transmission and reflection coefficients for different values of potential step height, for the first order splitting (there are similar results for the second order splitting). The ``calculated'' value is obtained from our numerical method while the ``theoretical'' value is obtained from Eq. (\ref{eq:trans_co}). }
\label{fig:res_trans}
\end{figure}

The height of the potential barrier is varied and the reflection (R) and transmission (T) coefficients are measured after a time of 0.22 a.u. which corresponds to 50000 time iterations. These coefficients are calculated as
\begin{eqnarray}
 R = \frac{\int_{\mathbb{R}^{+}} |\psi(x)|^{2}}{\int_{\mathbb{R}} |\psi(x)|^{2}} \;\; \mbox{and} \;\;
T = \frac{\int_{\mathbb{R}^{-}} |\psi(x)|^{2}}{\int_{\mathbb{R}} |\psi(x)|^{2}} . 
\end{eqnarray}
The results of the computation for the transmission and reflection coefficients are shown in Fig. \ref{fig:res_trans} for the first order splitting while a typical initial and reflected wave functions are shown in Fig. \ref{fig:klein_density}.

Relative differences between the analytical and numerical results are smaller than 0.5\% for all values of potential height. Moreover, the transmission coefficient is null for $V_{0} < 4.1 \times 10^{4}$ a.u., which is also in accordance with the analytical analysis.

\subsubsection{2-D}

\begin{figure}[htbp]
	\centering
	\includegraphics[width=0.5\textwidth]{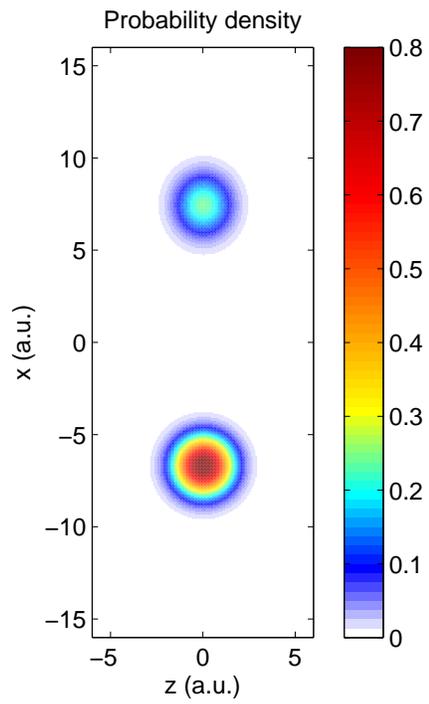}
	\caption{Final wave function at $t = 0.214$ a.u. after it scattered with the potential step. The transmitted wave function can be found in the region $x>0$ while the reflected part is in $x<0$. Note that the potential step is centred on $x=0$. }
	\label{fig:klein_2d}
\end{figure}

The same system is now analyzed in 2-D. Again, the initial state is a Gaussian wave packet prepared at $t=0$ given by
\begin{equation}
 \psi(t=0, x) = \mathcal{N} \begin{bmatrix} 1 \\ 0 \\ 0 \\ C \end{bmatrix} e^{ik_0 x} e^{- \frac{(x-x_{0})^{2} +(z-z_{0})^{2}}{4 (\Delta )^{2}}} \;\; \mbox{with} 
\end{equation}
with a given momentum $k_{0}=100.$ a.u.. Its width is $\Delta = 1.0$ a.u. in both $x$ and $z$ dimensions. The potential has a height $V_{0} = mc^{2} + 1.0 \times 10^{4}$ a.u. and a thickness of $L = 0.0025$ a.u.. The domain boundaries are set at $x=\pm 15.0$ a.u. and $z=\pm 5.0$ a.u.. The result of the computation for the wave function density is shown in Fig. \ref{fig:klein_2d} at a time of $t=0.214$ a.u., after the wave packet has scattered with the potential barrier. In this figure, it is very easy to see the transmitted (for $x>0$) and reflected parts ($x<0$) of the wave function, in accordance with the 1-D analysis. The value of the transmission and reflection coefficients are 0.2407  and 0.7593 respectively, which is relatively close (a relative difference of less than 9\% for the transmission and 3\% for the reflection coefficients) to the analytical results which are 0.2621 and 0.7380 respectively.

To obtain this accuracy and to resolve the quasi-discontinuous potential barrier, it was necessary to use a very large mesh. The result plotted in Fig. \ref{fig:klein_2d} were obtained with $24576 \times 8192$ elements and a second order splitting (thus, a time step of $\delta t = 1.78 \times 10^{-5}$ a.u.). To evolve the wave function to the final time $t=0.214$ a.u., 12000 time iterations were necessary. With such a large mesh, the computation time was approximately $T_{\rm comp.} \approx 22$ hours on 256 processors\footnote{Again, this calculation was carried on the MAMMOUTH cluster described previously.}, showing the performance of our numerical method. A better accuracy would require a mesh with the same element size as in the 1-D case, i.e. around $a \approx 3 \times 10^{-4}$ (for the same $a$, it was verified that the 1-D and 2-D cases give approximately the same transmission and reflection coefficients).

%
%

\section{Conclusion}
\label{sec:conclu}

In this work, a numerical method for the solution of time-dependent Dirac equation in real space was analyzed and used in simple applications. One of its main features is that it does not suffer from the fermion-doubling problem, as demonstrated in our analysis. The method is based on a combination of the operator splitting and characteristic methods and thus, is relatively simple. This simplicity resulted in very good parallelization performance and in very efficient calculations. Its main drawback however is the relation between the time and space increment which has to be chosen as $c \delta t = K \delta x$. This condition limits the flexibility for the space grid size in certain circumstances: for instance, if the space variations of the wave function are small, it would be more efficient to use larger element size but still keeping a small timestep (the latter is often necessary because there is a term like $e^{-i mc^{2} t}$ in the Dirac equation solution which oscillates rapidly). This is possible with the method described in this paper only if some modifications to the numerical scheme are implemented. For instance, one could use a Lax-Wendroff scheme to solve Eqs. (\ref{eq:split1}) to (\ref{eq:split3}) instead of the method of characteristics, but this would introduce again the fermion doubling problem. Note that it is possible to analyze systems for which the space variation of the wave function is large and the time variation is small by choosing $K>1$.

We applied this method to study simple systems such as the Klein paradox in 1-D and 2-D, and the time evolution of Gaussian wave packets (other more complex systems are under investigation). The results obtained from our method were compared to analytical computations and showed a very good agreement. However, to obtain the required accuracy, very small meshes had to be utilized, especially for reproducing the phase of the wave function. One interesting avenue to circumvent these difficulties would be to consider higher order splitting (such as third of fourth order splittings) \cite{Bandrauk2006346}. Another interesting possibility is the use of the GPU architecture which can lead to significant improvement in computation time \cite{Bauke2011}. Work is under way to verify if these techniques would result in better computation performance.  

\section*{Acknowledgement}

We would like to thank Heiko Bauke and Szczepan Chelkowski for useful comments. Also, we acknowlege the support of Huizhong Lu concerning computational issues and for pointing out the reason for the superlinear scaling behaviour of our algorithm. 

\appendix


\section{Solution of the Dirac Equation in Each Dimension}
\label{sec:app1}

This Appendix describes how the solution of the Dirac equation in each dimension is obtained. We are interested in solving the following equation:
\begin{equation}
 i\partial_{t} \psi(t,\mathbf{x}) = -i c\alpha_{i} \partial_{i}   \psi(t,\mathbf{x})
\end{equation}
with an initial condition given by
\begin{equation}
 \psi(t_{0},\mathbf{x}) = \phi(\mathbf{x}) .
\end{equation}
Note here that the index $i=x,y,z$ is not summed and thus, each dimension is treated independently. In the first step, the matrix $\alpha_{i}$ is diagonalized to decouple the spinor components. This is performed by a similarity transformation as
\begin{equation}
 P_{i}^{\dagger}\alpha_{i}P_{i} = \Lambda_{i}
\end{equation}
where $P_{i}$ are unitary matrices and $\Lambda_{i} = \mathrm{Diag}[1,1,-1,-1] = \beta$ is a diagonal matrix. Starting with $\alpha_{i}$ in the Dirac representation, the explicit expression of the transformation matrix is
\begin{equation}
 P_{i} = \frac{1}{\sqrt{2}} \left(\beta + \alpha_{i} \right).
\end{equation}
The representation thus obtained is very closed to the Majorana representation, modulo the index of the matrix in Dirac representation that becomes $\beta$. 

The resulting Dirac equation is then 
\begin{equation}
 i\partial_{t} \psi '(t,\mathbf{x}) = -i c\beta \partial_{i}  \psi '(t,\mathbf{x})
\end{equation}
where we defined $\psi ' = P^{\dagger} \psi$. It is convenient here, for notational purposes, to split the four-spinor into two bi-spinors as $\psi' = (\varphi',\chi')^{\rm T}$ to get
\begin{eqnarray}
 i\partial_{t} \varphi '(t,\mathbf{x}) &=& -i c \partial_{i}  \varphi '(t,\mathbf{x}) \\ 
 i\partial_{t} \chi '(t,\mathbf{x}) &=& i c \partial_{i}    \chi '(t,\mathbf{x})
\end{eqnarray}
Therefore, the Dirac equation clearly becomes a set of four uncoupled first-order differential equations in this representation. Their solutions are well-known and can be obtained from the method of characteristics. They are given by
\begin{eqnarray}
\varphi_{1,2} '(t,\mathbf{x}) &=& \phi_{1,2}'(x_{i}-c \delta t,\vec{x}) , \\
\chi_{1,2}'(t,\mathbf{x}) &=& \phi_{3,4}'(x_{i}+c \delta t,\vec{x}) ,
\end{eqnarray}
along with the conditions $x\pm x_{0} = c \delta t$. The latter is the characteristics equation along which the partial differential equation becomes an ordinary differential equation. Here we also use a particular notation with a function argument $(t,x_{i} \pm ct, \vec{x})$, which means that the substitution $x_{i} \rightarrow x_{i} \pm ct$ is performed while the other coordinates $\vec{x} \equiv (x_{j},x_{k})$ (with $i \neq j \neq k$) are unchanged. Note that the initial conditions are related to the original representation as $\phi' = P^{\dagger} \phi$.

In the last step, the solution is transformed back to the original representation. The final result, after some manipulations, is given by
\begin{eqnarray}
 \psi(t,\mathbf{x}) &=& \frac{1}{2} \left\{ [\mathbb{I}_{4} + \alpha_{i}] \phi(x_{i}-c\delta t,\vec{x})   + [\mathbb{I}_{4} - \alpha_{i}] \phi(x_{i}+c \delta t,\vec{x})  \right\}.
\end{eqnarray}
This equation is used in the numerical method to evolve the wave function in time in alternate direction iteration.

\section{Solution of 2-D wave packet}
\label{app:sol_2D_WP}

In this Appendix, the analytical solution for the time evolution of 2-D free wave packet is computed. We consider an initial wave packet for a massive spin-up electron at rest. The initial wave function is given by Eq. (\ref{eq:psi_init_2DWP}) and its Fourier transform by 
\begin{equation}
 \widehat{\psi}(t=0,p_x,p_z) = 4 \pi \Delta^{2} \mathcal{N} 
\begin{bmatrix}
1 \\ 
0 \\ 
0 \\ 
0
\end{bmatrix}
e^{- \Delta^{2} (p_{x}^{2} + p_{z}^{2} ) }.
\end{equation}
The 2-D Dirac equation we want to solve is given by
\begin{equation}
 i\partial_{t} \widehat{\psi}(t,p_{x},p_{z}) = \left[ c\alpha_{x} p_{x} + c\alpha_{z} p_{z}  + \beta m c^{2} \right]\widehat{\psi}(t,p_{x},p_{z}),
\end{equation}
here expressed in Fourier space. The solution to this equation is then simply
\begin{eqnarray}
\widehat{\psi}(t,p_{x},p_{z}) &=& e^{ -ic \alpha_{x} p_{x}t - ic \alpha_{z} p_{z}t -i\beta mc^{2}t } \widehat{\psi}(0,p_{x},p_{z}) \\
&=& \left[ \mathbb{I}_{4} \cos \left( E t \right) -i \frac{c\alpha_{x} p_{x} + c\alpha_{z}p_{z} + \beta mc^{2}}{E} \sin \left(  E t \right) \right] \nonumber \\
&& \times \widehat{\psi}(0,p_{x},p_{z}) ,
\end{eqnarray} 
where $E = \sqrt{p_{x}^{2}c^{2} + p_{z}^{2}c^{2} + m^{2}c^{4}}$. This last equation can be Fourier transformed back to real space. Then, using polar coordinates ($p^{2} = p_{x}^{2} + p_{z}^{2}$) and properties of the Bessel functions, we get the solution as
\begin{eqnarray}
\label{eq:WP_an_2d_1}
 \psi_{1}(t,x,z) &=& 2 \mathcal{N} \Delta^{2} \int_{0}^{\infty} dp p e^{-\Delta^{2}p^{2}} J_{0}(pr)  \nonumber \\ 
&& \times \left [ \cos(Et) - i\frac{mc^{2}}{E}\sin(Et) \right]  \\
\label{eq:WP_an_2d_2}
 \psi_{2}(t,x,z) &=& 0 \\
\label{eq:WP_an_2d_3} 
\psi_{3}(t,x,z) &=& -2 \mathcal{N} \Delta^{2} \sin(\phi)\int_{0}^{\infty} dp p J_{1}(pr) \frac{cp}{E} \sin(Et) e^{-\Delta^{2}p^{2}} \\
\label{eq:WP_an_2d_4}
 \psi_{4}(t,x,z) &=& -2 \mathcal{N} \Delta^{2} \cos(\phi)\int_{0}^{\infty} dp p J_{1}(pr) \frac{cp}{E} \sin(Et) e^{-\Delta^{2}p^{2}} 
\end{eqnarray}
where $J_{n}(z)$ is the Bessel function of the first kind and $\phi = \arctan (z/x)$ is the polar angle in real space. There is no known solution for these integrals in the general case \cite{watson} but they can be found for two specific cases when the mass is zero ($m=0.0$): at $r=0$ and at $r = ct $. If $r=0$, we have
\begin{eqnarray}
 \psi_{1}(t,0,0) &=& \mathcal{N} \left [1 -    \sqrt{\pi} \frac{ct}{2 \Delta} e^{-\frac{(ct)^{2}}{4\Delta^2}} \mathrm{Erfi}\left( \frac{ct}{2 \Delta }\right) \right] \\
 \psi_{2}(t,0,0) &=& \psi_{3}(t,0,0) = \psi_{4}(t,0,0) = 0 
\end{eqnarray}
where $\mathrm{Erfi}(z)$ is the imaginary error function. 

If $r=ct$ we have
\begin{eqnarray}
 \psi_{1}(t,x,z) &=&  \mathcal{N} \; _{2}F_{2}\left( \frac{1}{4},\frac{3}{4} ; \frac{1}{2}, \frac{1}{2} ; -\frac{(ct)^2}{\Delta^2} \right) \\
 \psi_{2}(t,x,z) &=& 0 \\
 \psi_{3}(t,x,z) &=& -\mathcal{N} (ct)^{2} \sin(\phi) \; _{2}F_{2}\left( \frac{5}{4},\frac{7}{4} ; \frac{3}{2}, \frac{5}{2} ; -\frac{(ct)^2}{\Delta^2} \right) \\
 \psi_{4}(t,x,z) &=& -\mathcal{N} (ct)^{2} \cos(\phi) \; _{2}F_{2}\left( \frac{5}{4},\frac{7}{4} ; \frac{3}{2}, \frac{5}{2} ; -\frac{(ct)^2}{\Delta^2} \right) 
\end{eqnarray}
where $_{2}F_{2}$ is the generalized hypergeometric series. These two results will be used to validate our numerical method and to analyze the operator splitting. These two analytical solutions at the origin (at $(x,z)=(0,0)$) and at $x^{2} + z^2 = ct$ for the time evolution of a 2-D massless wave packet are useful mostly for the comparison with the numerical method and for validation purposes.

\section{Solution of 1-D wave packet}
\label{app:sol_1D_WP}

In this Appendix, the analytical solution for the time evolution of a 1-D free wave packet is computed.  This calculation follows closely the one for the 2-D wave packet in \ref{app:sol_2D_WP} and for this reason, only the main steps are presented. The initial wave function is given by Eq. (\ref{eq:trav_wave}) and its Fourier transform by 
\begin{equation}
 \widehat{\psi}(t=0,p_x) = 2 \sqrt{\pi} \Delta \mathcal{N} 
\begin{bmatrix}
1 \\ 
0 \\ 
0 \\ 
C
\end{bmatrix}
e^{- \Delta^{2} (p_{x}-k_{0})^{2}   }.
\end{equation}
The 1-D Dirac equation we want to solve is given by
\begin{equation}
 i\partial_{t} \widehat{\psi}(t,p_{x}) = \left[ c\alpha_{x} p_{x} + \beta m c^{2} \right]\widehat{\psi}(t,p_{x}),
\end{equation}
here expressed in Fourier space. The solution to this equation is then simply
\begin{eqnarray}
\widehat{\psi}(t,p_{x}) &=& \left[ \mathbb{I}_{4} \cos \left( E t \right) -i \frac{c\alpha_{x} p_{x} + \beta mc^{2}}{E} \sin \left(  E t \right) \right]  \widehat{\psi}(0,p_{x}) ,
\end{eqnarray} 
where $E = \sqrt{p_{x}^{2} c^{2} + m^{2}c^{4}}$. This last equation can be Fourier transformed back to real space and we get the solution as
\begin{eqnarray}
\label{eq:WP_an_1d_1_t}
 \psi_{1}(t,x) &=&  \mathcal{N} \frac{\Delta}{\sqrt{\pi}} \int_{-\infty}^{\infty} dp e^{-\Delta^{2}(p-k_0)^{2}} e^{ipx}  \nonumber \\ 
&& \times \left [ \cos(Et) - i\frac{mc^{2}}{E}\sin(Et) - i C\frac{cp}{E} \sin(Et)\right]  ,\\
\label{eq:WP_an_1d_2_t}
 \psi_{2}(t,x) &=& 0 ,\\
\label{eq:WP_an_1d_3_t}
 \psi_{3}(t,x) &=& 0 ,\\
\label{eq:WP_an_1d_4_t}
 \psi_{4}(t,x) &=&  \mathcal{N} \frac{\Delta}{\sqrt{\pi}} \int_{-\infty}^{\infty} dp e^{-\Delta^{2}(p-k_0)^{2}} e^{ipx}  \nonumber \\ 
&& \times \left [ C\cos(Et) + iC\frac{mc^{2}}{E}\sin(Et) - i \frac{cp}{E} \sin(Et)\right]  .
\end{eqnarray}
There are no analytical solutions to these integrals, so they have to be computed numerically. 

We can find the solution for an initial wave packet for a massive spin-up electron at rest by setting $k_0 = 0$ to obtain
\begin{eqnarray}
\label{eq:WP_an_1d_1}
 \psi_{1}(t,x) &=&  \mathcal{N} \frac{\Delta}{\sqrt{\pi}} \int_{-\infty}^{\infty} dp e^{-\Delta^{2}p^{2}} e^{ipx}  \nonumber \\ 
&& \times \left [ \cos(Et) - i\frac{mc^{2}}{E}\sin(Et) \right]  ,\\
\label{eq:WP_an_1d_2}
 \psi_{2}(t,x) &=& 0 ,\\
\label{eq:WP_an_1d_3}
 \psi_{3}(t,x) &=& 0 ,\\
\label{eq:WP_an_1d_4}
 \psi_{4}(t,x) &=& -i \mathcal{N} \frac{\Delta}{\sqrt{\pi}} \int_{-\infty}^{\infty} dp e^{-\Delta^{2}p^{2}} e^{ipx} \frac{cp}{E} \sin(Et) .
\end{eqnarray}

\bibliographystyle{elsarticle-num}
\bibliography{bibliography}

\end{document}